%% file: majun.tex
\documentclass[preprint2]{aastex}

\newcommand\beq{\begin{equation}}
\newcommand\eeq{\end{equation}}

\received{}
\accepted{}

\lefthead{M} \righthead{M81 Globular Clusters: Colors and
Metallicities}

\begin{document}

{\title{Color and Metallicity Distributions of M81 Globular
Clusters}

\author{
Jun Ma\altaffilmark{1}, Xu Zhou\altaffilmark{1}, Jiansheng
Chen\altaffilmark{1}, Zhenyu Wu\altaffilmark{1}, Yanbin
Yang\altaffilmark{1}, Zhaoji Jiang\altaffilmark{1}, Jianghua
Wu\altaffilmark{1}}

\altaffiltext{1}{National Astronomical Observatories,
Chinese Academy of Sciences, Beijing, 100012, P. R. China;
majun@vega.bac.pku.edu.cn}

\authoremail{majun@vega.bac.pku.edu.cn}

\begin{abstract}
In this paper we present catalogs of photometric and spectroscopic
data for M81 globular clusters (GCs). The catalogs include $B-$ and
$V-$ photometric and reddening data of 95 GCs, and spectroscopic
metallicities of 40 GCs in M81. Using these data, we make some
statistical correlations. The results show that the distributions
of intrinsic $B$ and $V$ colors and metallicities are bimodal,
with metallicity peaks at $\rm {[Fe/H]}\approx -1.45$ and $-0.53$, respectively
as has been demonstrated for our Milky Way and M31. The relation between spectroscopic
metallicity and intrinsic $B$ and $V$ color also exists as it does for the
Milky Way and M31.

\end{abstract}


\section{Introduction}

Great progress has been made in the past decade in our
understanding of globular cluster systems (GCSs) of galaxies,
especially the discovery that many galaxies possess two or more
distinct subpopulations of globular clusters (GCs)
\citep[e.g.,][and references therein]{west04}. GCs are fossils of
the earliest stages of galaxy formation, and are groups of
stellar populations with a single age and metallicity. Consequently, their
integrated properties, such as abundance and kinematics, can provide
us with valuable information about the nature and duration of the
formation of their parent galaxies \citep{bh00}. The
metallicity distribution of GCs is of particular importance in
deepening our knowledge of the dynamical and chemical evolution of
the parent galaxies. For example, the GCs of many elliptical
galaxies show multi-modal metallicity distributions, suggesting
that multiple star formation episodes occurred in these elliptical
galaxies in the past \citep{zepf93,bh00}.

\citet{patr99} presents a metallicity distribution of 133 Galactic
GCs that apparently shows two peaks (i.e., two distinct metal-poor
and metal-rich GC populations). The double-Gaussian can best fit
these two subpopulations, the mean metallicity values are $-1.59$
and $-0.55$ dex, respectively. Using the data for 247 GCs in M31,
\citet{bh00} studied the metallicity distribution, which is
asymmetric, implying the possibility of bimodality. Then the
applied KMM algorithm showed that the metallicity distribution is
really bimodal. \citet{perrett02} confirmed the conclusions of
\citet{bh00}.

M81 is the nearest large spiral outside the Local Group, and its
GCs are an important target of study. However, to date, there are
only a few papers that studied the GCs in M81. The main reason is
that at the distance of M81, the diameters of GCs are typically
comparable to the seeing disk, and it is difficult to recognize
GCs on the basis of image structure from ground-based
observations, except at sites with exceptionally good seeing. To
maximize the success rate of the GC candidate list for the ongoing
spectroscopic observations, \citet{pr95} used an extensive
database that included photometric, astrometric, and morphological
information on 3774 objects covering over a $>50~\rm{arcmin}$
diameter field centered on M81 to reveal 70 GC candidates within
11 kpc galactocentric radius. \citet{pbh95} then confirmed 25
$bona$ $fide$ M81 GCs from the spectroscopy of 82 bright GC
candidates in the M81 field. \citet{sbkhp02} presented
moderate-resolution spectroscopy for 16 GC candidates from the
list in \citet{pr95}, and confirmed these 16 candidates as $bona$
$fide$ GCs. They also obtained metallicities for 15 of the 16 GCs.
With the superior resolution of the $Hubble$ $Space$ $Telescope$
$(HST)$ , M81 is close enough for its clusters to be easily
resolved on the basis of image structure \citep{cft01}. Thus,
using the $B$, $V$, and $I$ bands of $HST$ Wide Field Planetary
Camera 2 (WFPC2), \citet{cft01} imaged eight fields covering a
total area of $\sim 40~\rm{arcmin}^{2}$. They reported 114 compact
star clusters in M81, 59 of which are GCs.

The outline of the paper is as follows.  Details of collecting GCs
are given in \S~2. In \S~3 we provide some statistical
relationships. The summary and discussion are presented in \S~4.

\section{A Catalog of Photometric Data of Globular Clusters}

\subsection{Sample of Globular Clusters}

The sample of GCs of M81 in this paper is from
\citet{pbh95,sbkhp02,cft01}. \citet{pbh95} obtained spectra for 82
bright GC candidates in the M81 field and confirmed 25 $bona$
$fide$ GCs. \citet{sbkhp02} observed moderate-resolution
spectroscopy of 16 GC candidates selected from the candidate list
of \citet{pr95} and confirmed all as $bona$ $fide$ GCs.
\citet{cft01} discovered 114 compact star clusters (as mentioned
above), 59 of them confirmed GCs. Clusters Id50552, Is50696, and
Id50826 of \citet{pbh95} are numbers 87, 40 and 8 of
\citet{cft01}, and objects 13 and 15 of \citet{sbkhp02} are 1 and
7 of \citet{cft01}, respectively. Altogether, there are 95 GCs,
which are listed in Table 1 [col. (1) is the name of the GC, and
cols. (2) and (3) are the $V$-band magtitude and $(B-V)_0$ color,
respectively].

\subsection{Reddening}
In order to obtain intrinsic colors for the GCs, in addition to
the absolute magnitudes, the photometric data should be corrected
for reddening from the foreground extinction contribution of the
Milky Way and for the internal reddening due to varying optical
paths through the disk of M81. The total reddening in M81
(foreground plus M81 contribution) has been measured by a number
of authors \citep[e.g.,][]{fwm94,kong00}. We only mention here
that \citet{kong00} obtained the reddening maps of M81 based on
the images in 13 intermediate-band filters from 3800 to 10000 \AA.
To determine the metallicity, age, and reddening distributions for
M81, \citet{kong00} found the best match between the observed
colors and the predictions from the single stellar population
models of \citet{bc96}. A map of the interstellar reddening in a
substantial portion of M81 were obtained. We used the reddening
data of \citet{kong00}. For a few clusters that fall near the
edges of the images, \citet{kong00} did not present the reddening
data, we adopt the mean reddening value of 0.13 as \citet{cft01}
did. Since \citet{cft01} also used the reddening data of Kong et
al. to deredden their star clusters, we only deredden the GCs of
\citet{pbh95} and \citet{sbkhp02}. These local reddening values
are listed in column (4) of Table 1. For completeness, we also
list the reddening data for the GCs of \citet{cft01}.

\section{Properties of Globular Clusters}
\subsection{Luminosity Function and Color Magnitude Diagram}

We show in Figure 1 the luminosity functions for all the sample
GCs in M81. The magnitudes are corrected for extinction based on
the local $E_{(B-V)}$ value given in column (3) of Table 1 and the
optical Galactic extinction law with $R_v=3.1$ and a distance
modulus for M81 of 27.8 \citep{fwm94,cft01}. We can see that the
M81 GC luminosity function is not unimodal (without an apparent
peak). However, with a fainter completeness limit, the turnover
may appear.

\begin{figure}[htbp]
\epsscale{0.8} \hspace{0.0cm}\rotatebox{-90}{\plotone{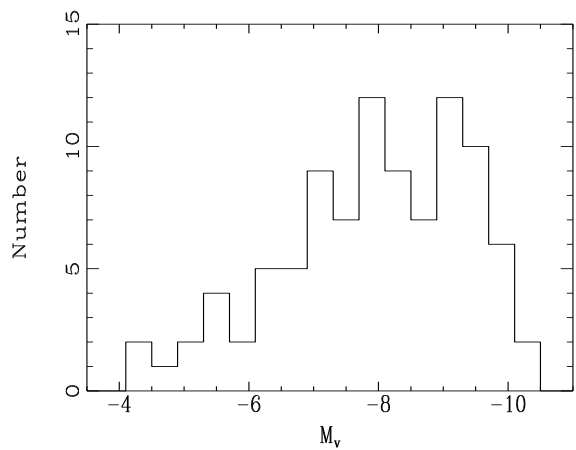}}
\caption{Luminosity function of the sample globular clusters.}
\label{fig:one}
\end{figure}

Figure 2 plots a $(B-V)_0$ versus $M_V$ color-magnitude diagram
for the M81 GCs. Colors have been dereddened by the local
$E_{(B-V)}$ value given in column (3) of Table 1. From Figure 2,
we note that there are some GCs that have $(B-V)_0$ greater than
1.0. By comparing with the $(B-V)_0$ of Galactic GCs \citep{hr79}
and M31 GCs \citep{bh00}, these M81 GCs have great $(B-V)_0$.
These red colors may result from enhanced foreground reddening
comparing with the applied $E_{(B-V)}$. In particular, there is a
GC \citep[96 of][]{cft01} having the $(B-V)_0=1.778$, which is too
high.

\begin{figure}[htbp]
\epsscale{0.8} \hspace{0.0cm}\rotatebox{-90}{\plotone{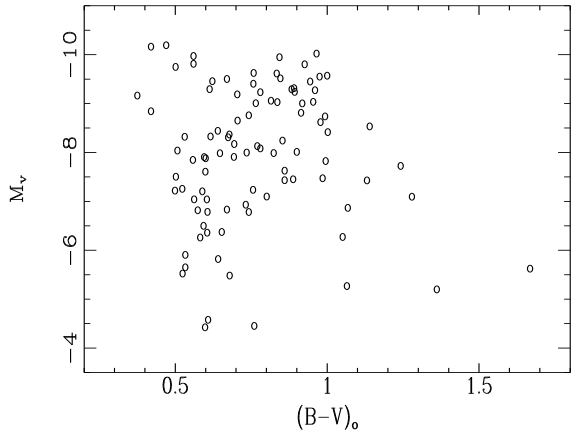}}
\caption{Color-magnitude diagrams of the sample globular
clusters.} \label{fig:two}
\end{figure}

Figure 3 shows the $(B-V)_0$ histogram for our sample globular
clusters, and two color peaks appear clearly. Using deep F555W and
F814W images from WFPC2 on board the $HST$, \citet{kb21} presented
the results of the GC systems of 28 elliptical galaxies, in which
the $V-I$ color distributions of at least $50\%$ of the sample
galaxies appear to be bimodal. In addition, from $HST$ imaging in
555W and F814W filters, \citet{larsen01} studied the GCs in 17
relatively nearly early-type galaxies and found that a sum of two
Gaussians provides a better fit to the observed color distribution
than a single Gaussian. To make quantitative statements about the
bimodality of $(B-V)_0$ in this paper, a KMM test \citep{abz94} is
applied to the data. This test uses a maximum likelihood method to
estimate the probability that the data distribution is better
modeled as a sum of two Gaussians than as a single Gaussian. Here
we use a homoscedastic test (i.e., the two Gaussians are assumed
to have the same dispersion). The $(B-V)_0$ of the two peaks, the
$P$-value, and the numbers of GCs assigned to each peak by the KMM
test are $(B-V)_0\approx 0.98$ and $0.66$, 0.071, and 67 and 27.
The $P$-value is in fact the probability that the data are drawn
from a single Gaussian distribution. Since the GC 96 of
\citet{cft01} has very high $(B-V)_0$ [$=1.778$], we do not
include it when conducting the KMM test. We note that the second
Gaussian peak looks like a terrible fit, which may be because the
data sample is not large enough. \citet{bh00} analyzed the
distribution of 10 intrinsic colors for the 221 M31 GCs and tested
the color distributions of these GCs for bimodality using the KMM
algorithm \citep{abz94}; the peaks of the $(B-V)_0$ are $0.83$ and
$0.68$. The peaks of the other intrinsic color distributions are
presented in Table 8 of \citet{bh00}. For the color distributions
of the Galactic GCs, \citet{bh00} only plotted the distribution of
$(V-I)_0$, and did not apply the KMM test. From Figure 14 of
\citet{bh00}, two clear peaks of distribution of $(V-I)_0$ can be
seen.

\begin{figure}[htbp]
\epsscale{0.8} \hspace{0.0cm}\rotatebox{-90}{\plotone{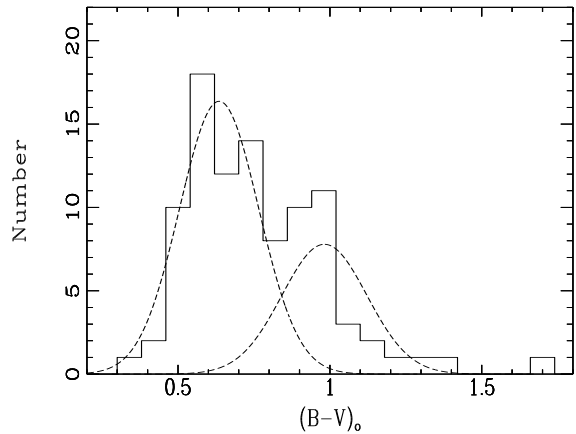}}
\caption{Color histogram distribution of the sample globular
clusters.} \label{fig:three}
\end{figure}

\subsection{Metallicity Distribution of the Sample Globular
Clusters Using Only with Spectroscopic Data}

\citet{pbh95} obtained the spectra for 25 M81 GCs with the Hydra
multifiber positioner and bench spectrograph on the KPNO 4 m
telescope. \citet{sbkhp02} then observed 16 M81 GCs with the Low
Resolution Imaging Spectrograph on the Keck I telescope and
presented the moderate-rosolution spectroscopy for these GCs, in
addition to the metallicities for 15 of the 16 GCs. There is one
GC, number 12 of \citet{sbkhp02}, for which the metallicity was
not obtained, since a transient phenomenon was occurring in this
GC at the time of the observation. Thus, there are a total of 40
GCs for which we have the metallicities, which are listed in Table
2. However, some of \citet{pbh95} sample GC metallicities have
large uncertainties, which can create uncertainties in the
metallicity distributions. Therefore, Figure 4 only shows the
metallicity diagram of 26 M81 GCs with metallicity uncertainties
smaller than 1.0 dex. Figure 4 does not show two peaks, and
perhaps because the number of sample GCs are too small. In order
to enlarge the number of GCs, we use metallicities estimated from
colors by the color-metallicity relation.

\begin{figure}[htbp]
\epsscale{0.8} \hspace{0.0cm}\rotatebox{-90}{\plotone{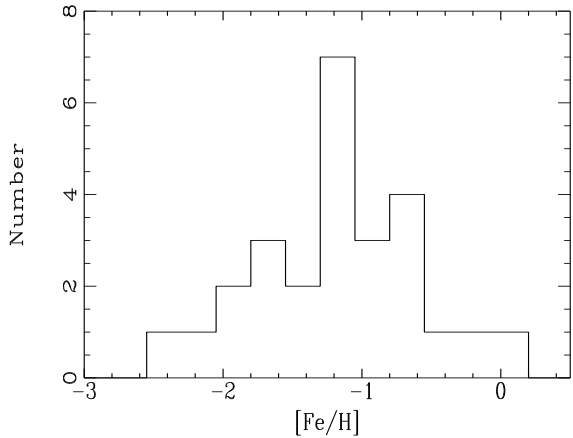}}
\caption{Metallicity diagram for the globular clusters only with
spectroscopic data.} \label{fig:four}
\end{figure}

\subsection{Color-Metallicity Correlation}

Originally, \citet{bh90} derived the correlation between the IR
colors and metallicity of 23 low-reddening Galactic GCs.
\citet{kb98} studied the GCs of NGC 1399 using
moderate-resolution, high signal-to-noise ratio spectroscopy and
found that $V-I$ and metallicity are well correlated. To determine
the reddening and use the color to predict metallicity,
\citet{bh00} performed linear regressions of the color-metallicity
for M31 and Galactic GCs.

We present below the relation between intrinsic color in the $B$
and $V$ bands and metallicity for 26 M81 GCs for which we have
spectroscopic metallicities with metallicity uncertainties smaller
than 1.0 dex. We do an ordinary least-squares fit:

\begin{equation}
{\rm [Fe/H]}=a(B-V)_0+b
\end{equation}

Figure 5 shows this fit. The fit results are $a=2.41\pm0.46$ and
$b=-3.08\pm0.37$, and the linear correlation coefficient $r=0.73$.

\citet{couture90} established the intrinsic $B-V$
color-metallicity relation for 65 Galactic GCs, using the data
from \citet{reed88} and \citet{zinn85}, for which the foreground
reddening is low ($E_{B-V}\leq 0.4$) and reasonably well known:

\begin{equation}
(B-V)_0=0.200{\rm [Fe/H]}+0.971
\end{equation}

\citet{bh00} also presented the intrinsic $B-V$ color-metallicity
relation for 88 Galactic GCs with $E_{B-V}<0.5$ (the database of
Galactic GC parameters were from the 1999 June version of the
\citet{harris96}):

\begin{equation}
(B-V)_0=0.159\pm 0.011{\rm [Fe/H]}+0.92\pm 0.02
\end{equation}

In order to make comparisons with the intrinsic $B-V$
color-metallicity relation for the Galactic GCs, we establish this
relation using the data from the 26 M81 GCs used in Figure 5:

\begin{equation}
(B-V)_0=0.222\pm 0.042{\rm [Fe/H]}+1.05\pm 0.05
\end{equation}

We can see that the intrinsic $B-V$ color-metallicity relation
between the Galactic GCs and the M81 GCs is consistent by
comparing equations (2), (3) and (4).

\begin{figure}[htbp]
\epsscale{0.8} \hspace{0.0cm}\rotatebox{-90}{\plotone{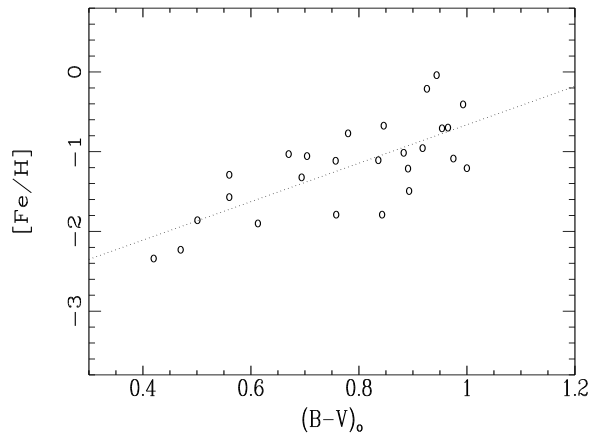}}
\caption{Color in $BV$ bands vs. spectroscopic metallicities for
M81 globular clusters.} \label{fig:five}
\end{figure}

\subsection{Metallicity Distribution of All the Sample Globular
Clusters}

We use the above color-metallicity correlation to derive
metallicities for the sample GCs that do not have spectroscopic
metallicities or that have metallicity uncertainties larger than 1
dex. We plot metallicity distribution in Figure 6, including the
metallicities derived using the above color-metallicity
correlation. To make quantitative statements about bimodality, we
also apply a KMM test \citep{abz94} to the data. The metallicities
of the two peaks, the $P$-value, and the numbers of clusters
assigned to each peak by the KMM test are $\rm {[Fe/H]}\approx
-1.45$ and $-0.54$, 0.024, and 74 and 20, respectively. Since the
GC 96 of \citet{cft01} has very high $(B-V)_0$ ($(B-V)_0=1.778$),
and the metallicity obtained using the color-metallicity
correlation is too rich (0.95 dex), we do not include it when
performing ing the KMM test.

\begin{figure}[htbp]
\epsscale{0.8} \hspace{0.0cm}\rotatebox{-90}{\plotone{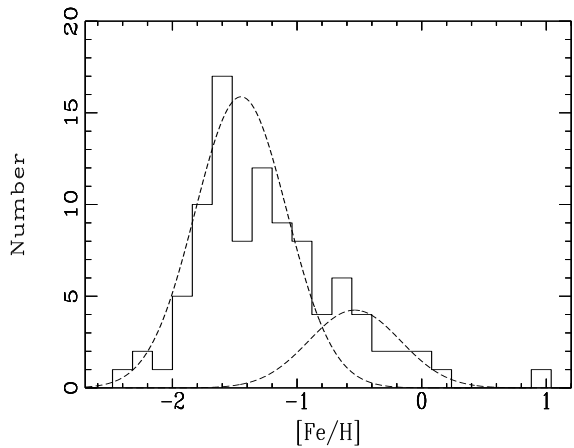}}
\caption{Metallicity diagram for all the globular clusters.}
\label{fig:six}
\end{figure}

\section{SUMMARY AND DISCUSSION}

In this paper, we present catalogs of photometric and
spectroscopic data and statistical correlations for M81 GCs. Using
the KMM test, we find that the distribution of intrinsic $B$ and
$V$ colors and metallicities are bimodal. The correlation between
spectroscopic metallicity and intrinsic $B$ and $V$ color also
exists as it does for the Milky Way and M31.

\acknowledgments We would like to thank the anonymous referee for his/her
insightful comments and suggestions that improved this paper very much.
This work has been supported by the Chinese
National Key Basic Research Science Foundation (NKBRSF
TG199075402) and in part by the Chinese National Science
Foundation, No. 10473012.

\clearpage
\input{table1.tex}
\clearpage
\input{table2.tex}
\end{document}

%% file: table1.tex
\begin{deluxetable}{cccc}
\tablenum{1}
\tablewidth{0pt}
\tablewidth{0pt}
\tablecaption{Globular cluster sample and properties}
\tablehead{\colhead{ID\tablenotemark{a}}&\colhead{V (mag)}&
\colhead{$B-V$ (mag)}&\colhead{$E_{B-V}$ (mag)}}
\tablecolumns{4}
\startdata
         Id30244 &  19.76 &   0.77 &  0.130\\
         Is40083 &  18.39 &   0.69 &  0.130\\
         Is40165 &  18.23 &   0.69 &  0.130\\
         Is40181 &  18.93 &   1.09 &  0.130\\
         Is50037 &  18.04 &   0.55 &  0.130\\
         Is50225 &  18.43 &   0.97 &  0.026\\
         Is50233 &  19.18 &   0.89 &  0.125\\
         Is50286 &  20.16 &   0.89 &  0.216\\
         Id50357 &  19.67 &   1.27 &  0.130\\
         Is50394 &  19.24 &   0.57 &  0.195\\
         Id50401 &  19.93 &   1.22 &  0.242\\
         Id50415 &  19.24 &   0.85 &  0.237\\
         Id50785 &  19.08 &   0.86 &  0.238\\
         Is50861 &  19.69 &   0.88 &  0.175\\
         Is50886 &  18.06 &   0.91 &  0.067\\
         Id50960 &  18.49 &   0.86 &  0.102\\
         Is51027 &  19.36 &   0.55 &  0.130\\
         Is60045 &  18.70 &   0.80 &  0.130\\
         Id70319 &  20.77 &   0.99 &  0.130\\
         Id70349 &  20.12 &   0.91 &  0.130\\
         Is80172 &  18.97 &   0.91 &  0.130\\
         Is90103 &  18.01 &   0.60 &  0.130\\
         SBKHP1 &  18.54 &   1.10 &  0.100\\
         SBKHP2 &  18.97 &   1.02 &  0.066\\
         SBKHP3 &  18.35 &   1.04 &  0.114\\
         SBKHP4 &  19.24 &   1.05 &  0.057\\
         SBKHP5 &  18.45 &   1.04 &  0.065\\
         SBKHP6 &  18.80 &   0.97 &  0.077\\
         SBKHP7 &  19.05 &   1.00 &  0.082\\
         SBKHP8 &  18.01 &   1.04 &  0.075\\
         SBKHP9 &  18.76 &   0.98 &  0.089\\
        SBKHP10 &  20.08 &   0.84 &  0.146\\
        SBKHP11 &  18.59 &   0.82 &  0.063\\
        SBKHP12 &  18.70 &   1.00 &  0.166\\
        SBKHP14 &  19.03 &   0.92 &  0.084\\
        SBKHP16 &  18.73 &   0.99 &  0.144\\
         CFT1 & 19.170 $\pm$  0.006 & 0.884 $\pm$  0.020 &  0.180\\
         CFT5 & 19.826 $\pm$  0.008 & 1.184 $\pm$  0.029 &  0.270\\
         CFT6 & 20.263 $\pm$  0.010 & 0.813 $\pm$  0.035 &  0.120\\
         CFT7 & 18.785 $\pm$  0.004 & 0.973 $\pm$  0.015 &  0.090\\
         CFT8 & 20.191 $\pm$  0.008 & 1.030 $\pm$  0.025 &  0.130\\
        CFT15 & 22.462 $\pm$  0.065 & 0.584 $\pm$  0.130 &  0.060\\
        CFT16 & 22.258 $\pm$  0.056 & 0.731 $\pm$  0.149 &  0.090\\
        CFT20 & 21.340 $\pm$  0.026 & 0.790 $\pm$  0.085 &  0.120\\
        CFT21 & 21.860 $\pm$  0.034 & 0.793 $\pm$  0.109 &  0.140\\
        CFT22 & 20.668 $\pm$  0.020 & 1.095 $\pm$  0.105 &  0.110\\
        CFT28 & 23.688 $\pm$  0.131 & 0.870 $\pm$  0.163 &  0.110\\
        CFT30 & 20.649 $\pm$  0.015 & 1.094 $\pm$  0.053 &  0.270\\
        CFT31 & 22.086 $\pm$  0.053 & 1.231 $\pm$  0.208 &  0.180\\
        CFT32 & 20.478 $\pm$  0.018 & 1.372 $\pm$  0.085 &  0.130\\
        CFT34 & 21.029 $\pm$  0.023 & 0.906 $\pm$  0.072 &  0.150\\
        CFT37 & 19.882 $\pm$  0.009 & 0.661 $\pm$  0.013 &  0.130\\
        CFT38 & 19.835 $\pm$  0.006 & 0.808 $\pm$  0.012 &  0.130\\
        CFT39 & 19.319 $\pm$  0.005 & 0.832 $\pm$  0.009 &  0.090\\
        CFT40 & 18.298 $\pm$  0.002 & 0.581 $\pm$  0.004 &  0.080\\
        CFT41 & 19.721 $\pm$  0.006 & 0.696 $\pm$  0.009 &  0.080\\
        CFT42 & 20.033 $\pm$  0.011 & 0.717 $\pm$  0.014 &  0.070\\
        CFT43 & 21.657 $\pm$  0.048 & 0.675 $\pm$  0.057 &  0.070\\
        CFT44 & 21.325 $\pm$  0.027 & 0.706 $\pm$  0.051 &  0.100\\
        CFT45 & 20.534 $\pm$  0.015 & 0.709 $\pm$  0.018 &  0.110\\
        CFT46 & 20.905 $\pm$  0.016 & 0.689 $\pm$  0.021 &  0.100\\
        CFT49 & 21.911 $\pm$  0.054 & 0.702 $\pm$  0.079 &  0.120\\
        CFT51 & 20.681 $\pm$  0.014 & 1.231 $\pm$  0.035 &  0.100\\
        CFT53 & 20.947 $\pm$  0.016 & 0.653 $\pm$  0.024 &  0.130\\
        CFT56 & 20.259 $\pm$  0.010 & 0.710 $\pm$  0.014 &  0.110\\
        CFT58 & 23.065 $\pm$  0.090 & 1.511 $\pm$  0.142 &  0.150\\
        CFT62 & 19.051 $\pm$  0.004 & 0.915 $\pm$  0.007 &  0.100\\
        CFT63 & 21.230 $\pm$  0.026 & 0.654 $\pm$  0.260 &  0.080\\
        CFT65 & 20.668 $\pm$  0.027 & 0.622 $\pm$  0.033 &  0.120\\
        CFT66 & 20.133 $\pm$  0.017 & 0.627 $\pm$  0.032 &  0.120\\
        CFT67 & 21.209 $\pm$  0.035 & 0.842 $\pm$  0.038 &  0.110\\
        CFT68 & 21.070 $\pm$  0.020 & 0.704 $\pm$  0.025 &  0.100\\
        CFT74 & 21.038 $\pm$  0.019 & 0.652 $\pm$  0.025 &  0.090\\
        CFT75 & 20.514 $\pm$  0.011 & 0.795 $\pm$  0.019 &  0.200\\
        CFT76 & 20.890 $\pm$  0.014 & 0.599 $\pm$  0.019 &  0.100\\
        CFT80 & 22.521 $\pm$  0.036 & 0.653 $\pm$  0.063 &  0.120\\
        CFT83 & 23.686 $\pm$  0.061 & 0.698 $\pm$  0.117 &  0.100\\
        CFT85 & 23.532 $\pm$  0.046 & 0.708 $\pm$  0.084 &  0.100\\
        CFT87 & 19.787 $\pm$  0.008 & 1.132 $\pm$  0.036 &  0.130\\
        CFT90 & 23.307 $\pm$  0.106 & 1.315 $\pm$  0.718 &  0.250\\
        CFT96 & 22.517 $\pm$  0.047 & 1.778 $\pm$  0.328 &  0.110\\
        CFT97 & 19.980 $\pm$  0.008 & 0.870 $\pm$  0.023 &  0.100\\
       CFT101 & 22.627 $\pm$  0.132 & 0.779 $\pm$  0.160 &  0.100\\
       CFT102 & 21.015 $\pm$  0.027 & 1.379 $\pm$  0.273 &  0.100\\
       CFT103 & 21.703 $\pm$  0.035 & 0.723 $\pm$  0.109 &  0.130\\
       CFT104 & 20.205 $\pm$  0.012 & 0.865 $\pm$  0.034 &  0.130\\
       CFT105 & 19.867 $\pm$  0.011 & 0.953 $\pm$  0.031 &  0.100\\
       CFT106 & 20.596 $\pm$  0.027 & 0.968 $\pm$  0.061 &  0.080\\
       CFT108 & 20.544 $\pm$  0.027 & 0.980 $\pm$  0.051 &  0.120\\
       CFT109 & 21.274 $\pm$  0.056 & 1.178 $\pm$  0.097 &  0.110\\
       CFT110 & 21.328 $\pm$  0.075 & 0.842 $\pm$  0.089 &  0.100\\
       CFT111 & 21.011 $\pm$  0.031 & 0.901 $\pm$  0.062 &  0.100\\
       CFT112 & 22.236 $\pm$  0.111 & 0.643 $\pm$  0.106 &  0.110\\
       CFT113 & 20.192 $\pm$  0.014 & 1.065 $\pm$  0.058 &  0.070\\
       CFT114 & 20.262 $\pm$  0.020 & 0.658 $\pm$  0.066 &  0.100\\
\enddata
\tablenotetext{a}{SBKHP identifications are from
\citet{sbkhp02}; CFT identifications are from \citet{cft01};
The others are from \citet{pbh95}.}
\end{deluxetable}

%% file: table2.tex
\begin{deluxetable}{cccc}
\tablenum{2} \tablewidth{0pt} \tablecaption{Metallicities for some
sample globular clusters} \tablehead{\colhead{ID}&\colhead{${\rm
[Fe/H]}$}&\colhead{ID}&\colhead{${\rm [Fe/H]}$}} \tablecolumns{6}
\startdata
         Id30244 &  $-1.76\pm 1.78$ & Is60045 &  $-1.03\pm 0.97$\\
         Is40083 &  $-1.29\pm 0.80$ & Id70319 &  $-2.31\pm 1.69$\\
         Is40165 &  $-1.57\pm 0.43$ & Id70349 &  $-2.41\pm 1.15$\\
         Is40181 &  $ 0.64\pm 1.43$ & Is80172 &  $-0.77\pm 0.68$\\
         Is50037 &  $-2.34\pm 0.83$ & Is90103 &  $-2.23\pm 0.99$\\
         Is50225 &  $-0.04\pm 0.59$ & SBKHP1 &  $-1.207\pm 0.369$\\
         Is50233 &  $-1.75\pm 1.02$ & SBKHP2 &  $-0.707\pm 0.167$\\
         Is50286 &  $-0.04\pm 1.85$ & SBKHP3 &  $-0.211\pm 0.193$\\
         Id50357 &  $-3.62\pm 2.97$ & SBKHP4 &  $-0.407\pm 0.088$\\
         Is50394 &  $-1.50\pm 1.29$ & SBKHP5 &  $-1.086\pm 0.091$\\
         Id50401 &  $-0.04\pm 1.00$ & SBKHP6 &  $-1.493\pm 0.206$\\
         Id50415 &  $-1.90\pm 0.71$ & SBKHP7 &  $-0.955\pm 0.098$\\
         Id50552 &  $ 0.98\pm 2.04$ & SBKHP8 &  $-0.698\pm 0.058$\\
         Is50696 &  $-1.86\pm 0.50$ & SBKHP9 &  $-1.212\pm 0.133$\\
         Id50785 &  $-0.72\pm 1.17$ & SBKHP10 &  $-1.322\pm 0.356$\\
         Id50826 &  $-1.46\pm 1.11$ & SBKHP11 &  $-1.114\pm 0.409$\\
         Is50861 &  $-1.71\pm 1.00$ & SBKHP13 &  $-1.055\pm 0.062$\\
         Is50886 &  $-1.79\pm 0.87$ & SBKHP14 &  $-1.107\pm 0.074$\\
         Id50960 &  $-1.79\pm 0.64$ & SBKHP15 &  $-1.014\pm 0.713$\\
         Is51027 &  $-2.47\pm 1.01$ & SBKHP16 &  $-0.674\pm 0.044$\\
\enddata
\end{deluxetable}

%% file: majun.bbl
\begin{thebibliography}{}
\bibitem[Ashman, Bird, \& Zepf(1994)]{abz94} Ashman, K. A., Bird,
C. M., \& Zepf, S. E. 1994, \aj, 108, 2348

\bibitem[Barmby et al.(2000)]{bh00} Barmby, P., Huchra, J., Brodie, J.,
Forbes, D., Schroder, L., \& Grillmair, C. 2000, \aj, 119, 727

\bibitem[Brodie \& Huchra(1990)]{bh90} Brodie, J. P., \& Huchra,
J. P. 1990, \apj, 362, 503

\bibitem[Bruzual \& Charlot(1996)]{bc96} Bruzual, G., \& Charlot, S.
1996, unpublished

\bibitem[Chandar, Ford, \& Tsvetanov(2001)]{cft01} Chandar, R., Ford, H. C.,
\& Tsvetanov, Z. 2001, \aj, 122, 1330

\bibitem[$\rm {C\hat{o}t\acute{e}}$(1999)]{patr99} $\rm {C\hat{o}t\acute{e}}$, P.
1999, \aj, 118, 406

\bibitem[Couture, Harris, \& Allwright(1990)]{couture90} Couture, J., Harris, W. E., \&
Allwright, J. W. B. 1990, \apjs, 73, 671

\bibitem[Freedman, Wilson, \& Madore(1994)]{fwm94} Freedman, W. L.,
Wilson, C. D., \& Madore, B. F. 1994,
\apj, 427, 628

\bibitem[Harris \& Racine(1979)]{hr79} Harris, W. E., \& Racine, R.
1979, \araa, 17, 241

\bibitem[Harris(1996)] {harris96} Harris, W. H. 1996, \aj, 112,
1487

\bibitem[Kong et al.(2000)]{kong00} Kong, X., et al., 2000, \aj, 119, 2745

\bibitem[Kissler-Patig et al.(1998)]{kb98} Kissler-Patig, M.,
Brodie, J. P., Schroder, L. L., Forbes, D. A., Grillmair, C. J.,
\& Huchra, J. P. 1998, \aj, 115, 105

\bibitem[Kundu \& Whitmore(2001)]{kb21} Kundu, A., \& Whitmore, B.
C. 2001, \aj, 121, 2974

\bibitem[Larsen et al.(2001)]{larsen01} Larsen, S. S., et al.,
2001, \aj, 121, 2974

\bibitem[Perelmuter, Brodie, \& Huchra(1995)] {pbh95} Perelmuter,
J. M., Brodie, J. P., \& Huchra, J. 1995, \aj, 110, 620

\bibitem[Perrett et al.(2002)] {perrett02} Perrett, K. M.. et al.,
2002, \aj, 123, 2490

\bibitem[Perelmuter \& Racine(1995)] {pr95} Perelmuter, J. M., \&
Racine, R. 1995, \aj, 109, 1055

\bibitem[Reed, Hesser, \& Shawl(1988)] {reed88} Reed, B. C., Hesser, J. E., \& Shawl, S. J.
1988, \pasp, 100, 545

\bibitem[Schroder et al.(2002)] {sbkhp02} Schroder, L. L., Brodie, J. P.,
Kissler-Patig, M., Huchra, J. P., Phillips, A. C. 2002, \aj, 123,
2473

\bibitem[West et al.(2004)] {west04} West, M. J., $\rm {C\hat{o}t\acute{e}}$,
P., Marzke, R. O., \& Jordan, A. 2004, \nat, 427, 31

\bibitem[Zepf \& Ashman(1993)] {zepf93} Zepf, S. E., \& Ashman, K. A.
1993, \mnras, 264, 611

\bibitem[Zinn(1985)] {zinn85} Zinn, R. 1985, \apj, 293, 424

\end{thebibliography}
